\documentclass{ws-mpla}
\usepackage[super]{cite}
\usepackage{graphicx}
\begin{document}

\markboth{P. P. Abrantes, D. Szilard, F. S. S. Rosa {\normalfont \&} C. Farina}
{Resonance Energy Transfer at Percolation Transition}

\catchline{}{}{}{}{}

\title{RESONANCE ENERGY TRANSFER AT PERCOLATION TRANSITION}

\author{\footnotesize P. P. ABRANTES$^{\dagger}$, D. SZILARD, F. S. S. ROSA  and C. FARINA}
\address{Instituto de F\'isica, Universidade Federal do Rio de Janeiro, Rio de Janeiro, Brazil \\
$^{\dagger}$patricia@if.ufrj.br}

\maketitle

\pub{Received (Day Month Year)}{Revised (Day Month Year)}

\begin{abstract}
We compute the resonance energy transfer (RET) in  a system composed of two quantum emitters near a host dielectric matrix in which metallic inclusions are inserted until the medium undergoes a dielectric-metal transition at percolation. 
We show that there is no peak in the RET rate at percolation, in contrast to what happens with the spontaneous emission rate of an emitter near the same critical medium. 
This result suggests that RET does not strongly correlate with the local density of states.

\end{abstract}

\ccode{PACS Nos.: 32.50.+d, 34.35.+a, 42.50.-p, 42.50.Pq}


\section{\label{sec:Introduction}Introduction}	

Resonance energy transfer (RET) is a very common phenomenon in nature and constitutes an important mechanism by which an excited quantum emitter (donor) transfers its energy to a neighboring one (acceptor) in the ground state \cite{GovorovFRET}. A remarkable application is the photosynthesis process, in which chlorophyll molecules absorb light and transfer this excitation to other neighboring chlorophyll molecules. Energy transfer has been extensively discussed and allows a wide range of applications such as photovoltaics \cite{Chanyawadee2009}, luminiscence \cite{Baldo2010} and sensing \cite{Schuler2002}. Further details on the history of energy transfer can be found in Ref.~\refcite{GovorovFRET}. 
Modern advances in nanophotonics, plasmonics and metamaterials, as well as the development of new technologies, paved the way for new mechanisms  of understanding and controlling the radiation-matter interaction at micro and nanoscales. In principle, such a control may be achieved by changing the profile of field modes - the so-called local density of states (LDOS) - by introducing macroscopic bodies in the vicinities of the emitters.  However, results in the literature about the interdependence between the RET rate and the LDOS have led researchers to distinct conclusions (see Ref.~\refcite{Wubs2016} and references therein). Motivated by this controversy, we investigate the RET rate of a pair of two-level quantum emitters in the vicinity of a semi-infinite medium composed by a dielectric host matrix in which metallic inclusions are inserted until this medium undergoes a dielectric-metal phase transition (at percolation). This system is an appropriate scenario for this analysis since it has been shown 
that the LDOS presents a peak close to the percolation.\cite{Szilard2016} This paper is organized as follows. In Sec.~\ref{subsec:Methodology} we present the methodology. In Sec.~\ref{sec:Results} we present our main results and discussions while Sec.~\ref{sec:Conclusions} is left for conclusions.

\section{\label{subsec:Methodology}Methodology}	

\begin{figure}[ph]
\vspace*{-10pt}
\centerline{\includegraphics[width=2.7in]{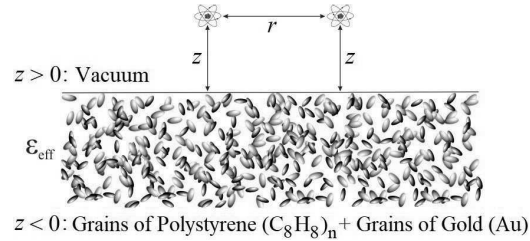}}
\caption{A pair of two-level emitters at a distance $z$ of a half-space composed of metallic (gold) inclusions embedded in a dielectric (polystyrene) host matrix. \protect\label{EmissoresMeio}}
\end{figure}

Figure~\ref{EmissoresMeio} shows a pair of two-level quantum emitters, $A$ and $B$, separated by a distance $r = |{\bf r}_B - {\bf r}_A|$, one of them in the excited state and the other one in the ground state, both at the same distance $z$ from a semi-infinite medium, composed by randomly distributed spherical metallic inclusions (with permittivity $\varepsilon_i$) embedded in a dielectric host matrix (with  permittivity $\varepsilon_{hm}$). The RET rate $\Gamma^{(m)}$ of emitters $A$ and $B$  normalized by the RET rate in free space $\Gamma^{(0)}$ can be written as \cite{Marocico2009}
\begin{equation}
\frac{\Gamma^{(m)}}{\Gamma^{(0)}} = \frac{\big| {\bf d}_B \cdot \mathbb{G} ({\bf r}_B, {\bf r}_A, \omega_0) \cdot {\bf d}_A \big|^2}{\big| {\bf d}_B \cdot \mathbb{G}^{(0)} ({\bf r}_B, {\bf r}_A, \omega_0) \cdot {\bf d}_A \big|^2} \, ,
\label{RETRate}
\end{equation}

\noindent where $\omega_0 = k_0 c$ is the transition frequency of the emitters, ${\bf d}_A$ and ${\bf d}_B$ are their transition electric dipole moments and $\mathbb{G} ({\bf r}, {\bf r}', \omega)$ and $\mathbb{G}^{(0)} ({\bf r}, {\bf r}', \omega)$ are the electromagnetic Green dyadics of the system and in free space respectively. The Green dyadic satisfies \cite{Novotny}
\begin{equation}
\left[ \nabla \! \times \! \nabla \! \times \, - \epsilon (\omega, {\bf r}) \frac{\omega^2}{c^2} \right] \mathbb{G} ({\bf r}, {\bf r}', \omega) = - \delta ({\bf r} - {\bf r}') \mathbb{I} 
\label{EMGDHelmoltz}
\end{equation}
(plus appropriate boundary conditions) and $\epsilon (\omega,{\bf r})$ is the electric permittivity of the medium. It is convenient to write the Green dyadic as ${\mathbb{G} ({\bf r}_B, {\bf r}_A, \omega) = \mathbb{G}^{(0)} ({\bf r}_B, {\bf r}_A, \omega) + \mathbb{G}^{(S)} ({\bf r}_B, {\bf r}_A, \omega)}$. The contribution $\mathbb{G}^{(S)} ({\bf r}_B, {\bf r}_A, \omega)$ takes into account the scattering due to the neighboring objects.~\cite{Novotny}
For simplicity, we consider both emitters with the transition dipole moments oriented along the $z$-axis, so that we only need the $zz$-component of the tensors in Eq.~(\ref{RETRate}). These equations depend explicitly on the Fresnel coefficient $r^{\textrm{TM,TM}}$ that can be expressed as \cite{Novotny} $r^{\textrm{TM,TM}} = (\varepsilon_e k_{0z} - k_{1z})(\varepsilon_e k_{0z} + k_{1z})^{-1}$, where $k_{1z} = \sqrt{\varepsilon_e k_0^2 - k_\parallel^2}$ and $\varepsilon_e$ is the effective dielectric constant of the critical medium to be computed by using Bruggeman effective-medium theory (BEMT) \cite{Choy}.  According to BEMT, an increase in the value of the volume filling factor $f$ means an increase in the ratio between the volume of metallic inclusions (gold in our case) and dielectric inclusions (polystyrene in our case). The expressions of the dielectric functions $\varepsilon_i$ and $\varepsilon_{hm}$ are
\begin{equation}
\varepsilon_i (\omega) = 1 - \frac{\omega_{pi}^2}{\omega^2 + i \gamma_i \omega} \;\;\;\;\; \textrm{and} \;\;\;\;\; \varepsilon_{hm} (\omega) = 1 + \sum_j \frac{ \omega_{pj}^2}{\omega_{Rj}^2 - \omega^2 - i \omega \Gamma_j} \,,
\end{equation}
\noindent where $\omega_{pi}$ ($\omega_{pj}$) and $\gamma_i$ ($\Gamma_j$) are, respectively, the plasma frequency (oscillating strengths) and the inverse of the relaxation time(s) of the metallic inclusions (host medium). The values of these parameters for gold and polystyrene were extracted from Refs.~\refcite{Ordal1985,Hough1980}.

\section{\label{sec:Results}Results and Discussions}

\begin{figure}[h]
\vspace*{-8pt}
\centerline{\includegraphics[width=5.0in]{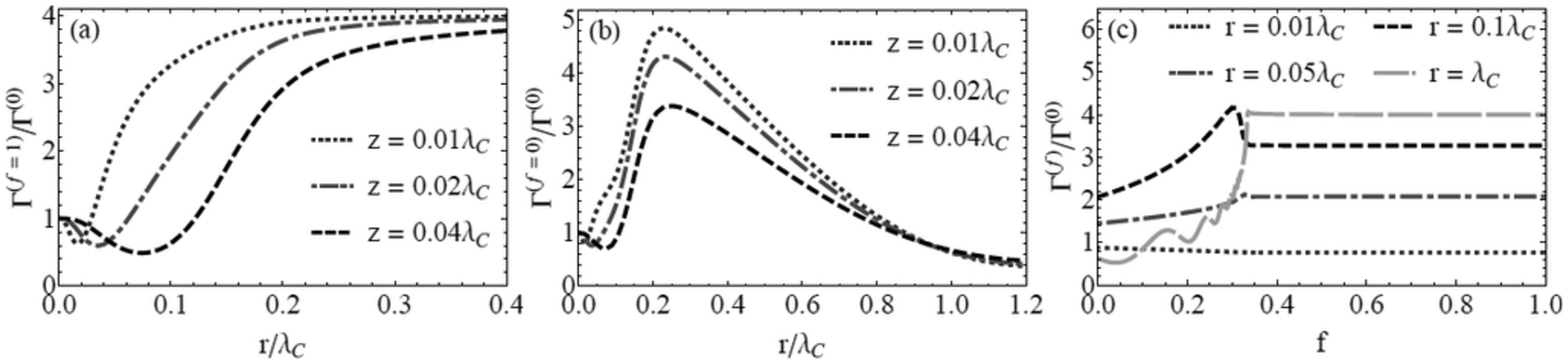}}
\caption{Normalized RET rate as a function of $r$ for three values of $z$ with the medium being gold (a) and polystyrene (b) 
	and as a function of the filling factor $f$ for three values of $r$ with $z = 0.01 \lambda_C$ (c). \protect\label{Graficos}}
\end{figure}

We start investigating the influence of different homogeneous media on the RET rate, one being composed of gold (filling factor $f = 1$) and the other of polystyrene ($f = 0$). The emitters are  cesium atoms with transition wavelength $\lambda_C = 450 \mu$m. In both Figs.~\ref{Graficos}(a) and \ref{Graficos}(b), the normalized RET rate is calculated as a function of the distance $r$ between the emitters for three different distances $z$ from them to the semi-infinite medium. In the first graph, the medium is made of gold and $\Gamma^{(f = 1)}/\Gamma^{(0)} \longrightarrow 1$ for $r/\lambda_C \longrightarrow 0$, as expected, since the closer the emitters are to each other, the less important is the influence of the medium. As $r$ is increased, the influence of the medium becomes more relevant. In the limit $r/\lambda_C \longrightarrow \infty$, we get $\Gamma^{(f = 1)}/\Gamma^{(0)} \longrightarrow 4$, a result that  can be understood in terms of images. For $z \longrightarrow 0$, the effects of the dipoles and their images are added, enhancing two times the energy transfer rate for each atom. The curve that most readily approaches the value four is the one corresponding to the smallest value of $z$.

In Fig.~\ref{Graficos}(b) the medium is polystyrene and once again $\Gamma^{(f = 0)}/\Gamma^{(0)} \longrightarrow 1$ for $r/\lambda_C \longrightarrow 0$. Unlike the previous case, as we increase $r/\lambda_C$ the normalized RET rate increases until it reaches a maximum and then steadily decreases. This was expected since the medium is a dielectric and the interpretation by image method is not so clear as it is for metals. Finally, returning to the main goal of this work, namely, the investigation of the dependence of RET on the LDOS, we show in Fig.~\ref{Graficos}(c) the normalized rate as a function of the filling factor $f$. Particularly, we are interested in the behavior close to percolation, where the LDOS has a peak \cite{Szilard2016}. At the critical value $f_c = 1/3$ for spherical inclusions, the medium undergoes a dielectric-metal transition, thereby exhibiting a dramatic change in its electrical and optical properties \cite{Kort-Kamp2014}. For very short distances between the emitters ($r\ll\lambda_C$) the presence of the medium is negligible, as it is evident from the dotted curve  ($r = 0.01\lambda_C$). As we increase $r$  the medium starts being noticed by the emitters. In these cases, as $f$ is increased  by the insertion of golden spheres, the images of the dipoles start to be better formed and hence the normalized RET rate increases. When $f=f_c$, the medium undergoes a dielectric-metal transition  so that for $f>f_c$ there will be no substantial change in each dipole image, which explains the plateau in the dashed-dotted curve ($r=0.05\lambda_C$). If we keep on increasing $r$, the contribution from the propagating modes comes into play, giving rise to the oscillations exhibited by the curves for $r=0.1\lambda_C$ (dashed line) and $r=\lambda_C$ (large dashed line). But in all of them we have plateaus for $f>f_c$, and no discernible peak close to the percolation.	

\section{\label{sec:Conclusions}Conclusions and Final Remarks}	

We studied the RET rate of a pair of two-level quantum emitters in the vicinity of a semi-infinite medium composed by randomly dispersed spherical golden inclusions in a polystyrene host matrix. This is an appropriate scenario to study the dependence of RET on the LDOS, since the latter has a pronounced peak close to the dielectric-metal transition (at percolation) \cite{Szilard2016}. As we have shown, RET rate has a negligible dependence on the LDOS.

\section*{Acknowledgments}

We thank R. M. Souza for discussions and CNPq and FAPERJ for financial support.

\end{document}